\documentclass[preprint]{article}
\usepackage{authblk}
\usepackage{graphicx}
\usepackage{dcolumn}
\usepackage{bm}
\usepackage{amsmath}
\usepackage{amssymb}
\usepackage{multirow}
\usepackage{caption}
\usepackage{subcaption}
\usepackage{epstopdf}
\usepackage{hyperref}

\begin{document}
\title{\bf Reconstructing cosmic expansion in $f(R, G)$ gravity using a log-periodic deceleration model}
\author[]{Amit Samaddar\thanks{samaddaramit4@gmail.com}}
\author[]{S. Surendra Singh\thanks{ssuren.mu@gmail.com}}
\affil[]{Department of Mathematics, National Institute of Technology Manipur, Imphal-795004,India.}

\maketitle
 
 \textbf{Abstract}: We study the late-time cosmology in $f(R, G)=R+\alpha R^{2}+\beta e^{\gamma G}$, using a logarithmic parametrization of the deceleration parameter $q(z)=q_{0}+q_{1}sin[log(1+z)]$. The Hubble parameter $H(z)$ is reconstructed and model parameters are constrained via MCMC analysis using CC ($31$), BAO ($26$) and Pantheon+SHOES ($1701$) datasets. Our results yield a Hubble constant in the range $H_0 = 71.7$--$72.8$ km/s/Mpc, consistent with late-time observations. The present deceleration parameter is found to be $q_{0}=-0.484$ to $-0.517$, while the evolution parameter $q_{1}\approx 1$, indicating increasing acceleration. The transition redshift shifts from $z_{tr}=0.879$ (CC) to $0.744$ (CC+BAO+Pantheon+SHOES), supporting a dynamic acceleration phase. The model reproduces early radiation behavior with $\omega (z>>1) \approx 0.33$ and predicts present-day values $\omega_{0} \approx -0.49$. Energy conditions NEC and DEC are satisfied, while SEC is violated at late times. The statefinder parameters $\{r_0, s_0\} = (0.866, 0.046)$ lie near the $\Lambda$CDM point. Estimated age of the Universe ranges from $13.01$ to $13.59$ Gyr. Thermodynamic analysis confirms consistency with the generalized second law. Overall, the model offers a viable and observationally consistent description of cosmic acceleration.

 
 \section{Introduction}\label{sec1}
 \hspace{0.5cm} Type Ia supernovae observations have delivered compelling evidence for the Universe's current accelerated expansion \cite{Riess98}. This phenomenon, often referred to as the dark energy era, stands as one of the most unexpected and intriguing discoveries in contemporary cosmology. Prior to its detection, only a few theorists had anticipated the possibility of such late-time acceleration. Since then, considerable research efforts have focused on explaining this phase of cosmic evolution \cite{S89}. Theoretical models proposed to account for the dark energy epoch span a wide spectrum—from scalar field-based frameworks like quintessence and other exotic fluids, to more geometric and structural modifications of gravity itself. Comprehensive reviews in the literature explore these diverse approaches, with modified gravity standing out as a particularly promising avenue \cite{Bamba12}-\cite{Li11}. Equally fascinating is the early Universe phase preceding the radiation-dominated era, known as inflation. During this epoch, the Universe is believed to have expanded exponentially within a very short time frame. Introduced in the early $1980$s, the inflationary paradigm addresses key issues in the classical Big Bang model, such as the horizon and flatness problems \cite{Guth81,Linde94}. Observational data from missions like Planck have significantly narrowed down the parameters of inflation, especially the spectral index and the tensor-to-scalar ratio. However, despite these successes, direct observational confirmation of inflation is still lacking \cite{Linde83}. As an alternative, bouncing cosmology has been proposed, suggesting a Universe that undergoes a contraction followed by an expansion, potentially offering a unified framework for both early- and late-time cosmic dynamics.
 
 Across a broad range of theoretical developments, modified gravity has emerged as a compelling framework capable of simultaneously describing both the early-time inflationary period and the current phase of cosmic acceleration. A coherent description of these two epochs has been made possible by innovative research like that in \cite{Nojiri03}. The core idea of modified gravity involves extending Einstein’s theory to accommodate new physical effects that become significant at high curvature regimes or on cosmological scales. General Relativity (GR), though remarkably successful at small and intermediate scales, appears to struggle with observations in the strong-curvature or large-scale limits. To account for late-time cosmic acceleration within the standard Einstein–Hilbert framework, one must introduce exotic components like phantom energy. However, such models often lead to unphysical outcomes, such as the Big Rip singularity \cite{Olmo11}, in which the Universe ends in a catastrophic disintegration. In contrast, modified gravity circumvents the need for such problematic energy sources by introducing higher-order curvature terms directly into the gravitational action. Numerous models based on this approach have been explored, as reviewed in \cite{Nojiri17}-\cite{Cald03}. $f(R)$ gravity is a well-known variant, characterized by replacing the Ricci scalar with a broader function $f(R)$. A well-established example is the Starobinsky model \cite{AA80}, which has shown strong consistency with both local gravitational tests and cosmological observations. Another significant model is the Nojiri–Odintsov proposal \cite{Nojiri03}, which provides a unified mechanism for explaining both inflation and late-time acceleration within the $f(R)$ framework. Expanding beyond $f(R)$, other modified theories incorporate additional geometric invariants, such as the Gauss–Bonnet term $G$, as in $f(G)$ gravity. These approaches have been successfully employed to model both early and late-time cosmic behavior \cite{Nojiri06}-\cite{Li07}. Further expansions include theories that consider higher-order curvature terms like $R^{ij}R_{ij}$ and $R^{ijkl}R_{ijkl}$, broadening the scope of gravitational models \cite{Clifton06}-\cite{Cap04}. The study of combined models, specifically $f(R,G)$ gravity, has intensified due to their promise in meeting both observational constraints and theoretical requirements \cite{KB10}-\cite{Ataz14}.
 
 The deceleration parameter $q(z)$ plays a central role in characterizing the expansion history of the Universe. It directly reflects whether the Universe is accelerating $(q<0)$ or decelerating $(q>0)$, and its evolution provides key insights into the transition between these two phases. Observational evidence strongly supports a cosmic history in which the Universe evolved from an early decelerated expansion to a present-day accelerated one. Capturing this transition accurately is essential for understanding both structure formation and the nature of cosmic acceleration. To explore the dynamic behavior of cosmic expansion, researchers have proposed numerous parameterizations of $q(z)$. These include simple linear forms such as $q(z)=q_{0}+q_{1}z$ as well as more flexible frameworks like the CPL-type models and other rational or inverse-square redshift dependencies \cite{Riess04}-\cite{Gong06}. Each of these forms attempts to describe the expansion rate over a broad range of redshifts, from the matter-dominated era to the far future. However, many conventional parameterizations are either too simplistic or restricted to narrow redshift ranges, limiting their ability to capture potential features such as oscillations or fine-scale deviations in the expansion rate \cite{Nair12}-\cite{AB24}. In contrast, this work introduces and analyzes a novel form of the deceleration parameter given by: $q(z)=q_{0}+q_{1}sin[log(1+z)]$ which embeds a periodic dependence on the logarithmic redshift scale. This log-periodic structure is designed to reflect possible scale-dependent fluctuations or oscillatory behavior in the cosmic expansion rate, which could emerge from high-energy physics or effective gravitational corrections. The model naturally accommodates a smooth transition from deceleration to acceleration and remains well-behaved across the entire observable redshift range, including the far future limit $z \to -1$. Moreover, this parametrization is not tied to a specific dark energy model or scalar field description, making it a model-independent yet flexible approach for testing cosmological dynamics. Importantly, the oscillatory term allows for the possibility of recurring phases of acceleration and deceleration on a logarithmic scale, which may be connected to features in the effective equation of state or deviations from standard gravity at different epochs.
 
 The structure of the paper is as follows. In Section \ref{sec2}, we outline the theoretical background of $f(R,G)$ gravity and derive the corresponding field equations for the chosen model. Section \ref{sec3} introduces the log-periodic parametrization of the deceleration parameter and presents the observational datasets, along with the MCMC-based estimation of the model parameters. In section \ref{sec4}, we analyze the behavior of the deceleration parameter, energy density, pressure and the equation of state parameter. Section \ref{sec5} is dedicated to the evaluation of energy conditions to assess the physical viability of the model. In section \ref{sec6}, we apply the statefinder diagnostic to further characterize the model's deviation from $\Lambda$CDM. Section \ref{sec7} estimates the age of the Universe using the reconstructed Hubble function. Section \ref{sec8} investigates the thermodynamic properties of the model, including the evolution of temperature and entropy density. Finally, section \ref{sec9} concludes the paper with a summary of results and their cosmological implications.
\section{Mathematical foundations of $f(R,G)$ gravity}\label{sec2}
\hspace{0.5cm} Our approach starts with a generalized action that modifies the standard Einstein-Hilbert form by incorporating the Gauss–Bonnet invariant through a functional dependence \cite{De15}.
\begin{equation}\label{1}
S=\int \bigg[\frac{1}{2\kappa^{2}}f(R,G)+\mathcal{L}_{m}\bigg]\sqrt{g} d^{4}x,
\end{equation}
Here, $g$ denotes the determinant of the metric, $\mathcal{L}_{m}$ represents the matter Lagrangian and $\kappa^{2}= 8\pi G_{N}$, with $G_{N}$ being the gravitational constant. The Gauss-Bonnet term is specified as
\begin{equation}\label{2}
G\equiv R^{2}-4R_{\mu\nu}R^{\mu\nu}+R_{\mu\nu\delta\sigma}R^{\mu\nu\delta\sigma},
\end{equation}
where the Ricci tensor is denoted by $R_{\mu\nu}$ and the Riemann tensor by $R_{\mu\nu\delta\sigma}$. Variation of the gravitational action (\ref{1}) with respect to $g_{\mu\nu}$ leads to the following gravitational field equations
\begin{eqnarray}\label{3}
G_{\mu\nu}&=&\frac{1}{f_{R}}\bigg[\kappa^{2}T_{\mu\nu}+\frac{1}{2}g_{\mu\nu}(f(R,G)-Rf_{R})+\nabla_{\mu}\nabla_{\nu}f_{R}-g_{\mu\nu}\Box f_{R}+f_{G}\big(-2RR_{\mu\nu}+4R_{\mu\delta}R^{\delta}_{\nu}\\\nonumber
&&-2R^{\delta\sigma\eta}_{\mu}R_{\nu\delta\sigma\eta}+4g^{\delta\sigma}g^{\eta\psi}R_{\mu\delta\nu\eta}R_{\sigma\psi}\big)+2\big(\nabla_{\mu}\nabla_{\nu}f_{G}\big)R-2g_{\mu\nu}\big(\Box f_{G}\big)R_{\mu\nu}+4\big(\Box f_{G}\big)R_{\mu\nu}\\\nonumber
&& -4\big(\nabla_{\delta}\nabla_{\mu}f_{G}\big)R_{\nu}^{\delta}-4\big(\nabla_{\delta}\nabla_{\nu}f_{G}\big)R_{\mu}^{\delta}+4g_{\mu\nu}\big(\nabla_{\delta}\nabla_{\sigma}f_{G}\big)R^{\delta\sigma}
-4\big(\nabla_{\delta}\nabla_{\psi}f_{G}\big)R_{\mu\delta\nu\eta}g^{\delta\sigma}g^{\eta\psi}\bigg].
\end{eqnarray}
Here, $G_{\mu\nu}$ denotes the Einstein tensor, $\nabla_{\mu}$ is the covariant derivative compatible with the metric $g_{\mu\nu}$, $\Box\equiv g^{\mu\nu}\nabla_{\mu}\nabla_{\nu}$ is the d’Alembert operator and $T_{\mu\nu}$ corresponds to the energy-momentum tensor describing the matter content. The subsequent variables are specified as follows:
\begin{equation}\label{4}
f_{R}=\frac{\partial f(R,G)}{\partial R} \hspace{0.4cm} \text{and} \hspace{0.3cm} f_{G}=\frac{\partial f(R,G)}{\partial G}.
\end{equation}
This study employs the FLRW metric for a flat Universe $(\kappa=0)$, characterized by a time-dependent scale factor $a(t)$. The metric takes the form
\begin{equation}\label{5}
ds^{2}=-dt^{2}+a^{2}(t)\big(dx^{2}+dy^{2}+dz^{2}\big),
\end{equation}
The Hubble parameter is defined as $H=\frac{\dot{a}}{a}$, with the overdot indicating a derivative with respect to time $``t"$. Accordingly, the expressions for the Ricci scalar and Gauss-Bonnet invariant are
\begin{equation}\label{6}
R=6(\dot{H}+2H^{2}) \hspace{0.3cm} \text{and} \hspace{0.3cm} G=24H^{2}(\dot{H}+H^{2}).
\end{equation}
Assuming an isotropic perfect fluid, the energy-momentum tensor yields the Einstein equations and continuity equation as,
\begin{equation}\label{7}
T_{\mu\nu}=diag(-\rho,p,p,p).
\end{equation}
Here, $\rho$ is the energy density and $p$ represents the pressure. Inserting the expressions from equations (\ref{5}) and (\ref{6}) into the field equation (\ref{3}), the $f(R,G)$ gravity field equations are
\begin{equation}\label{8}
\kappa^{2}\rho=3H^{2}f_{R}-\bigg(\frac{Rf_{R}+Gf_{G}-f(R,G)}{2}\bigg)+3H\dot{f}_{R}+12H^{3}\dot{f}_{G},
\end{equation}
\begin{equation}\label{9}
\kappa^{2}p=-(2\dot{H}+3H^{2})f_{R}+\bigg(\frac{Rf_{R}+Gf_{G}-f(R,G)}{2}\bigg)-2H\dot{f}_{R}-\ddot{f}_{R}-8H(\dot{H}+H^{2})\dot{f}_{G}-4H^{2}\ddot{f}_{G}.
\end{equation}

To explore the effects of higher-order curvature corrections in the gravitational action, we consider a modified gravity model characterized by the function
\begin{equation}\label{10}
f(R,G)=R+\alpha R^{2}+\beta e^{\gamma G},
\end{equation}
where $\alpha$, $\beta$ and $\gamma$ are the model parameters. This formulation generalizes General Relativity through quadratic and exponential curvature terms, motivated by various theoretical considerations. The $R^{2}$ term is mainly inspired by the Starobinsky inflationary model, which successfully describes early-Universe inflation through quantum gravity corrections. This exponential Gauss-Bonnet term is motivated by string theory and higher-dimensional gravity, where it helps to explain the Universe's late-time acceleration. Non-linear functions of $G$ enable more complex cosmological behavior, potentially preventing singularities and unifying early inflation with late dark energy phases. In the $\alpha\rightarrow0$ and $\beta\rightarrow0$ limit, the model recovers GR, reverting to the $f(R,G)=R$ form of the Einstein-Hilbert action. Different versions of this model have shown promise in reproducing $\Lambda$CDM-like expansion, resolving singularities, and ensuring stable cosmic dynamics without exotic matter. For example, \cite{Fomin17} explored inflationary dynamics in Einstein-Gauss-Bonnet gravity, revealing that a nonminimally coupled scalar field permits exponential and power-law inflationary scenarios. \cite{Bamba14} examined bouncing cosmologies in modified Gauss-Bonnet gravity, showing that exponential scale factors can lead to bouncing behavior in reconstructed models. These investigations demonstrate the adaptability of exponential Gauss-Bonnet terms in modeling diverse cosmological scenarios. In \cite{Li07}, the authors studied $f(G)$ gravity models, highlighting their capacity to reproduce cosmic acceleration without dark energy by modifying the gravitational action. These studies provide insight into the motivations and implications of curvature corrections in gravitational theories, demonstrating their ability to address cosmological problems and remain consistent with data.
\section{Parametric form of $q(z)$ and observational constraints}\label{sec3}
\hspace{0.5cm} In order to explore the late-time cosmic expansion in a model-independent yet flexible way, we consider a phenomenological parametrization of the deceleration parameter given by
\begin{equation}\label{11}
q(z)=q_{0}+q_{1}sin\big(log[1+z]\big),
\end{equation}
The oscillatory behavior of this form provides a versatile framework for understanding cosmic acceleration, accommodating fluctuations and transitions without dark energy model assumptions. The logarithmic argument yields a well-behaved, non-divergent description of the phenomenon, applicable to both low and high redshifts. Moreover, such oscillatory forms can encapsulate scenarios where the Universe may have undergone multiple acceleration–deceleration transitions, which are consistent with some theoretical models of modified gravity and dynamical dark energy. This approach, combining sine and logarithmic functions, aims to capture non-monotonic behavior in $q(z)$, going beyond the standard assumptions of simple linear or CPL-type models. Building on earlier work \cite{EM08,SN05}, logarithmic parametrizations offer a useful framework for capturing scaling behavior at high redshifts. Oscillatory models have been investigated in various frameworks, including interacting dark energy and modified gravity, where they can produce oscillations in key cosmological parameters, as seen in \cite{Ellis09,Alam04}. The advantages of this deceleration parameter include its regularity at $z=0$, adaptability to diverse observational trends, and compatibility with complex expansion histories featuring multiple transitions. This approach provides a useful framework for comparing alternative gravitational theories and evaluating cosmological models using current observational data.

It is worth emphasizing that, due to the limited availability and high degree of uncertainty in observational data for $q(z)$, we are unable to properly constrain the deceleration parameter forms. We focus on constraining $H(z)$ by integrating the chosen $q(z)$ parametrizations, rather than directly analyzing $q(z)$, which is
\begin{equation}\label{12}
H(z)=H_{0}exp\bigg[\int_{0}^{z}\frac{1+q(z)}{1+z}dz\bigg],
\end{equation}
Here, $H_{0}$ is the Hubble constant and its value inferred from the fits to observational data. Substituting of equation (\ref{11}) into equation (\ref{12}) yields the expression for $H(z)$ as:
\begin{equation}\label{13}
H(z)=H_{0}(1+z)^{1+q_{0}}e^{q_{1}[1-cos(log[1+z])]},
\end{equation}
The model parameters $H_{0}$, $q_{0}$ and $q_{1}$ will be constrained using observational data in the following section.
\subsection{Determining $H_{0}$, $q_{0}$ and $q_{1}$ through observational data}\label{sec3.1}
\hspace{0.5cm} This section explains how we constrain the assumed $q(z)$ parametrization with observational data. To determine the best-fit parameters, we adopt a likelihood function of the form $\mathcal{L}\propto e^{-\frac{\chi^{2}}{2}}$, where the joint chi-square is the sum of individual contributions from different datasets: $\chi^{2}_{joint}=\chi^{2}_{CC}+\chi^{2}_{BAO}+\chi^{2}_{SN}$. To extract the most compatible parameter values, we confront the adopted functional forms with observational datasets, utilizing the Markov Chain Monte Carlo (MCMC) approach to reduce the joint chi-square statistic, $\chi^{2}_{joint}$, constructed from all relevant data sources \cite{AL02,WR96}. The parameter estimation is carried out under the assumption of uniform priors within physically reasonable intervals: $H_{0}\in[60,100]$, $q_{0}\in[-2,2]$ and $q_{1}\in[0,2]$.
\subsubsection{Cosmic chronometers}\label{sec3.1.1}
\hspace{0.5cm} Hubble parameter measurements offer a window into the Universe's expansion history, leveraging a range of observational approaches. Notably, the Cosmic Chronometer (CC) approach is recognized for its robustness and minimal model dependence. This approach relies on passively evolving galaxies—systems that have ceased significant star formation—identified through their unique spectral signatures and color indices \cite{Moresco16,AS25}. This approach has provided $31$ data points for the Hubble parameter, covering the redshift interval $0.07\leq z\leq2.41$ \cite{NM23}. $H(z)$ is estimated in the CC approach by applying the differential age evolution of galaxies via: $H(z)=-\frac{1}{1+z}\frac{dz}{dt}$, with $\frac{dz}{dt}$ signifying the rate of redshift variation, derived from galaxy age differences. The $\chi^{2}$ function is employed to evaluate the model's compatibility with the data and is expressed as:
\begin{equation}\label{14}
\chi^{2}_{CC}=\sum_{j=1}^{31}\frac{\big[H(z_{j},H_{0},q_{0},q_{1})-H_{obs}(z_{j})\big]^{2}}{\sigma^{2}(z_{j})},
\end{equation}
Here, $H(z_{j},H_{0},q_{0},q_{1})$ gives the expected value of the Hubble parameter at redshift $z_{j}$, based on the specified cosmological parameters $(H_{0}, q_{0}, q_{1})$. The observed Hubble parameter value at redshift $z_{j}$ is given by $H_{obs}(z_{j})$ and $\sigma^{2}(z_{j})$ indicates the associated error.
\subsubsection{BAO dataset}\label{sec3.1.2}
\hspace{0.5cm} Baryon Acoustic Oscillations (BAO) are a valuable cosmological tool, tracing the signatures of primordial sound waves in the Universe's large-scale structure. The observed oscillatory features stem from early Universe density perturbations within the tightly coupled baryon-photon fluid prior to the epoch of recombination. In this work, we place constraints on cosmological models by employing BAO data derived from galaxies, quasars \cite{DESI24} and Lyman-$\alpha$ forest tracers \cite{AG25}, as reported in the initial data release from the Dark Energy Spectroscopic Instrument (DESI) survey \cite{AGA25}. BAO analyses typically focus on dimensionless observational quantities, with the first being defined as follows:
\begin{equation}\label{15}
\frac{d_{M}(z)}{r_{d}}=\frac{D_{L}(z)}{r_{d}(1+z)},
\end{equation}
The transverse comoving distance is given by $d_{M}(z)$ and $r_{d}$ is the sound horizon at the drag epoch, sensitive to matter, baryon and relativistic degrees of freedom. We adopt the method from \cite{DJ05} to compute $r_{d}$. Additionally, we utilize the Hubble distance ratio:
\begin{equation}\label{16}
\frac{d_{H}(z)}{r_{d}}=\frac{c}{r_{d}H(z)},
\end{equation}
Here, $d_{H}(z)$ denotes the Hubble distance. We also examine the ratio of the volume-averaged angular diameter distance:
\begin{equation}\label{17}
\frac{d_{V}(z)}{r_{d}}=\frac{\big[zd_{M}^{2}(z)d_{H}(z)\big]^{\frac{1}{3}}}{r_{d}},
\end{equation}
where $d_{V}(z)$ represents the comoving distance measured as an average over volume. For BAO data, the $\chi^{2}$ function based on $\frac{d_{M}(z)}{r_{d}}$, $\frac{d_{H}(z)}{r_{d}}$ and $\frac{d_{V}(z)}{r_{d}}$ is given by the following expression:
\begin{equation}\label{18}
\chi^{2}_{BAO}=\sum_{j=1}^{N} \bigg[\frac{Y_{j}^{th}-Y_{j}^{obs}}{\sigma_{Y_{i}}}\bigg]^{2},
\end{equation}
\subsubsection{Pantheon+SHOES data}\label{sec3.1.3}
\hspace{0.5cm} Type Ia supernovae observations have been pivotal in demonstrating the Universe's accelerated expansion. Type Ia supernovae serve as luminous benchmarks, enabling accurate distance measurements and shedding light on cosmic acceleration. Two decades of progress in SNeIa observations have resulted in extensive and refined datasets, notably Union, Union $2.1$, JLA, Pantheon and the latest Pantheon$^{+}$ dataset \cite{Kowa08}-\cite{Scolnic18}. Our analysis was based on the Pantheon+SHOES dataset which is the largest and most precise compilation of SNe Ia data currently available. This compilation encompasses $1701$ light curves of $1550$ SNe Ia, probing redshifts from $z = 0.001$ to $z = 2.26$. The formula for the theoretical distance modulus $\mu_{j}^{th}$ is:
\begin{equation}\label{19}
\mu_{j}^{th}(z,\Theta)=5\; \text{log}_{10}d_{L}(z,\Theta)+25,
\end{equation}
in the set of parameter $\Theta=[H_{0},q_{0},q_{1}]$, the luminosity distance $d_{L}(z,\Theta)$ takes the form:
\begin{equation}\label{20}
d_{L}(z,\Theta)=c(1+z)\int_{0}^{z}\frac{d\tilde{z}}{H(\tilde{z})},
\end{equation}
Regarding the Pantheon+SHOES Dataset, the distance modulus's chi-square function $(\chi^{2})$ is
\begin{equation}\label{21}
\chi^{2}(z,\Theta)=\sum_{j=1}^{1701}\bigg(\frac{\mu_{obs}(z_{j})-\mu_{th}(z_{j},\Theta)}{\sigma_{\mu}(z_{j})}\bigg)^{2},
\end{equation}
in which $\mu_{obs}$ is the measured distance modulus and $\mu_{th}$ is the theoretically calculated distance modulus. $\sigma_{\mu}(z_{j})$ embodies the measurement uncertainty in $\mu_{obs}$.

\begin{figure}[hbt!]
  \centering
  \includegraphics[scale=0.33]{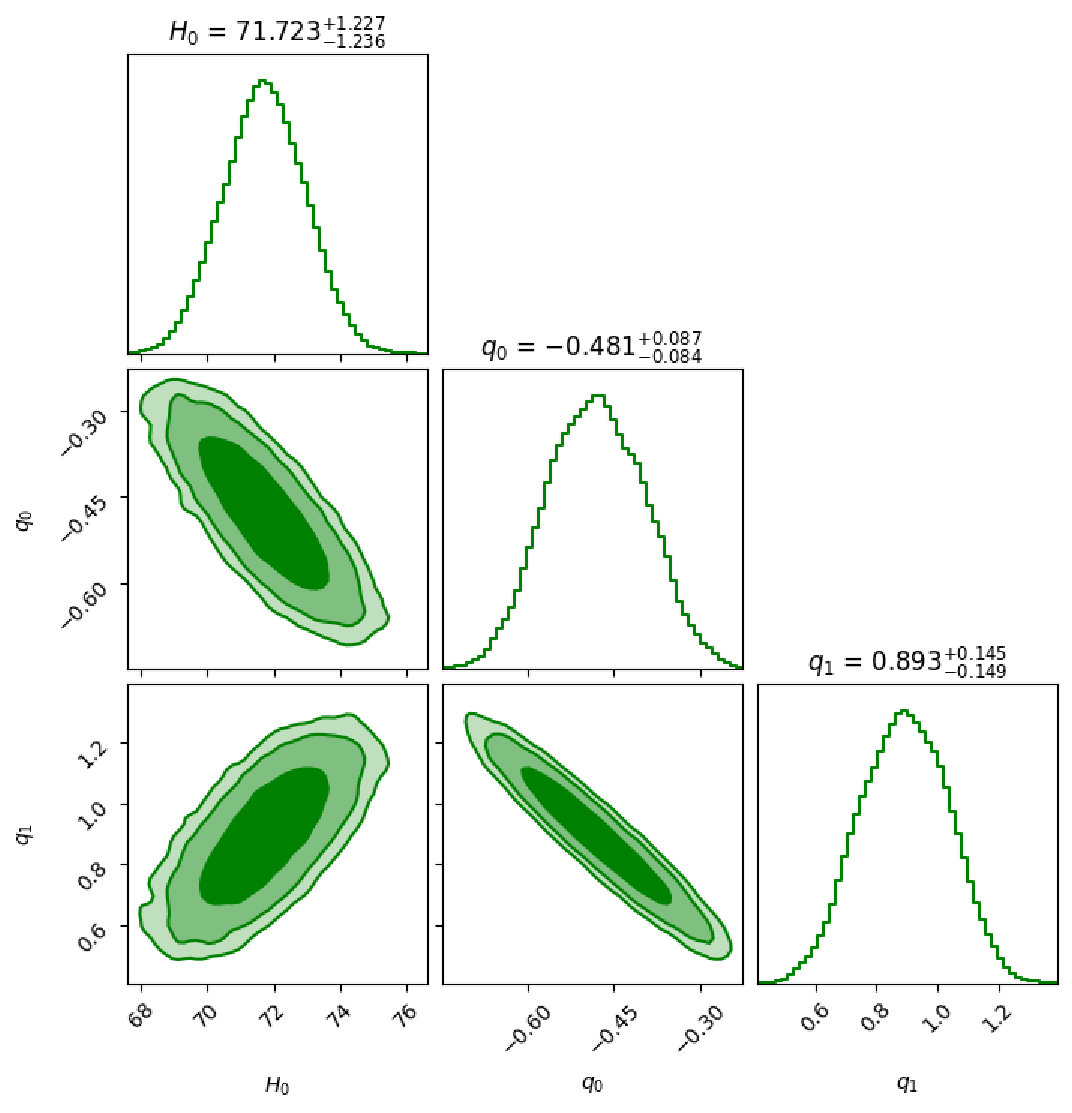}~~~
  \includegraphics[scale=0.33]{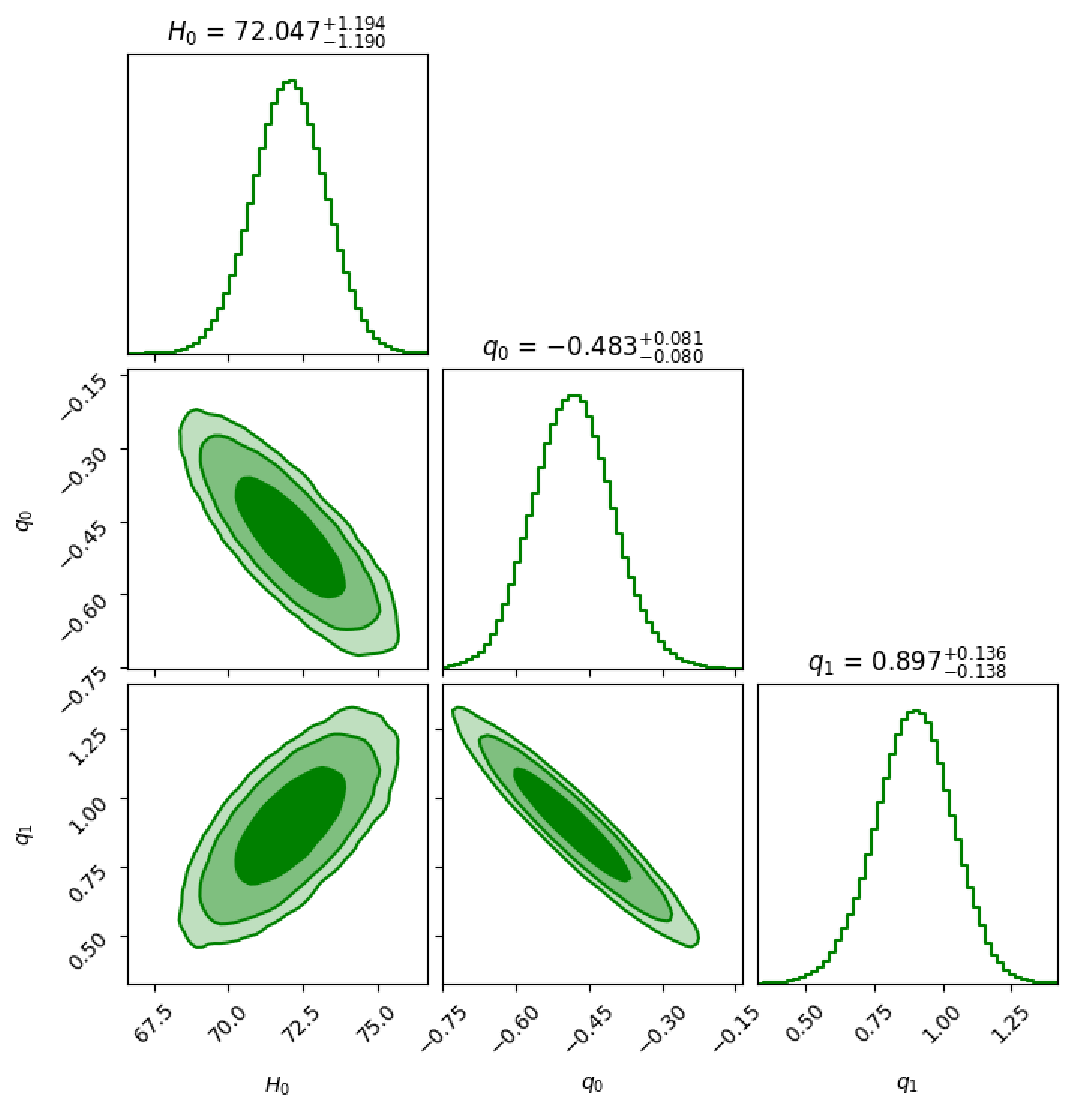}\\
  \includegraphics[scale=0.33]{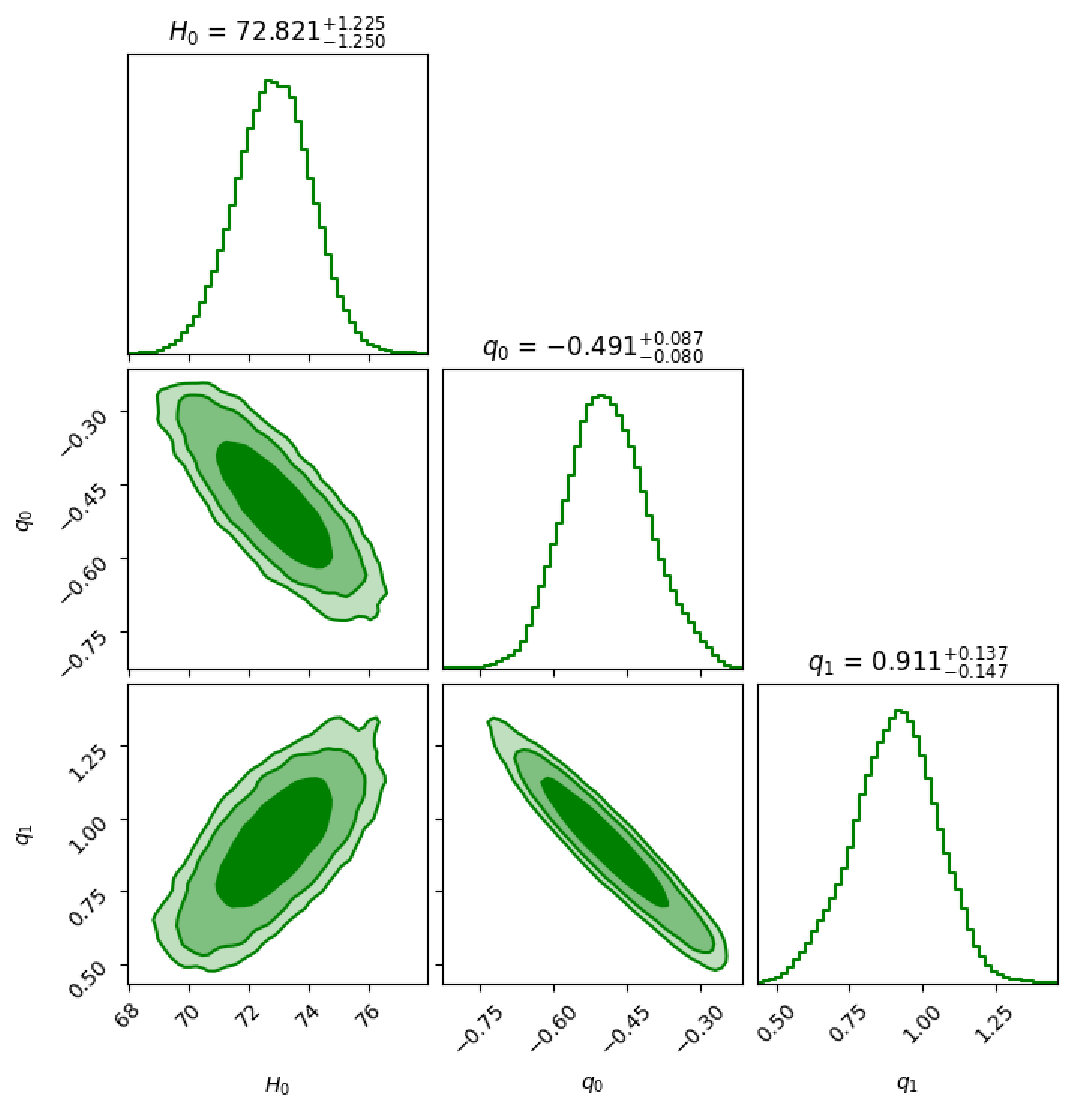}
  \caption{Confidence contour plots for the parameters $(H_{0}, q_{0}, q_{1})$ obtained from different combinations of datasets. The shaded regions denote the $1\sigma$ $(68.27\%)$, $2\sigma$ $(95.45\%)$ and $3\sigma$ $(99.73\%)$ confidence levels.}\label{fig:f1}
\end{figure}

\begin{figure}[hbt!]
  \centering
  \includegraphics[scale=0.3]{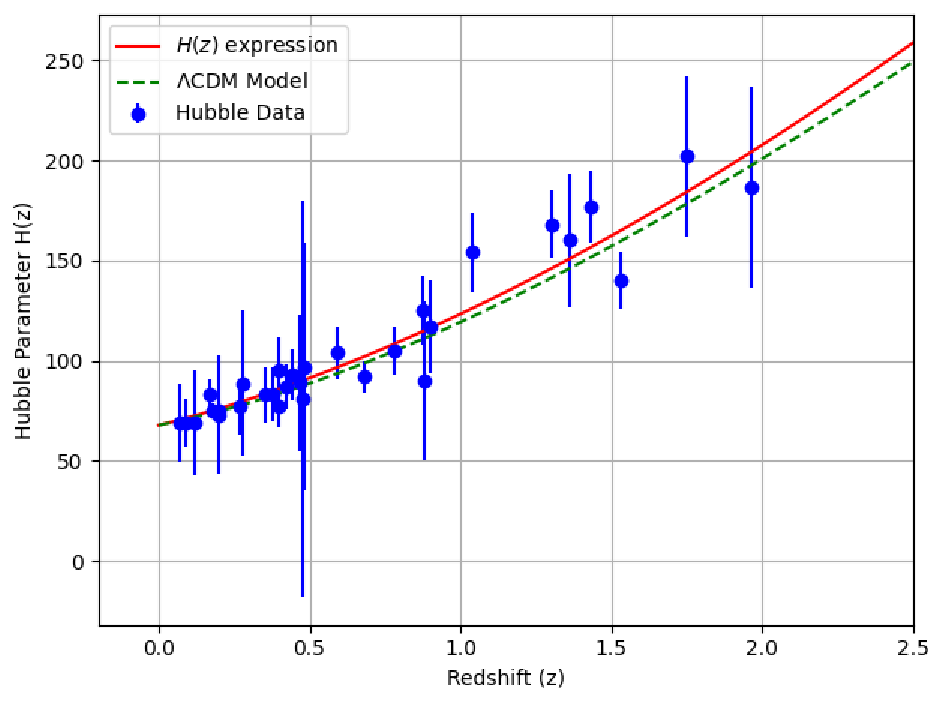}~
  \includegraphics[scale=0.33]{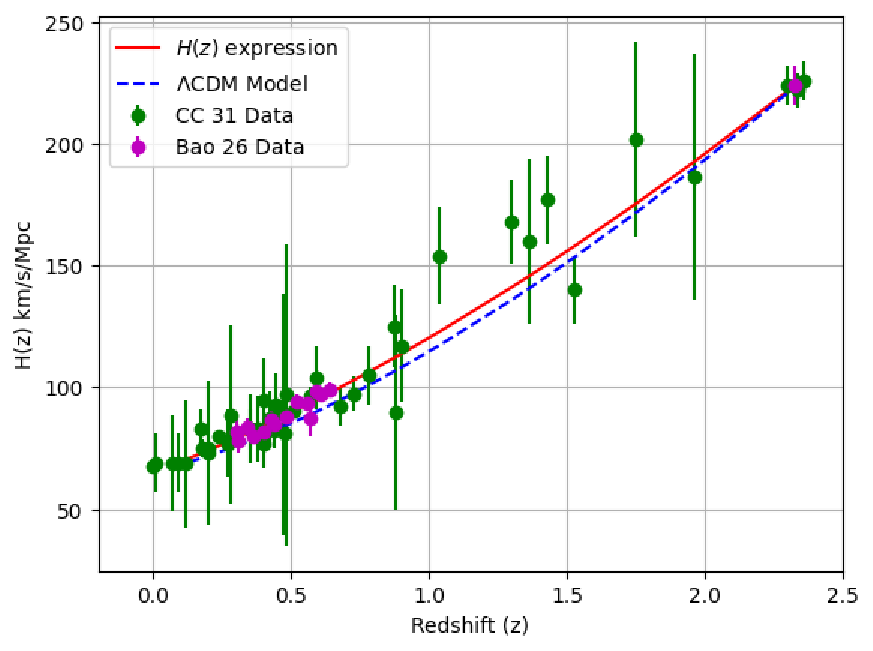}~~
  \includegraphics[scale=0.28]{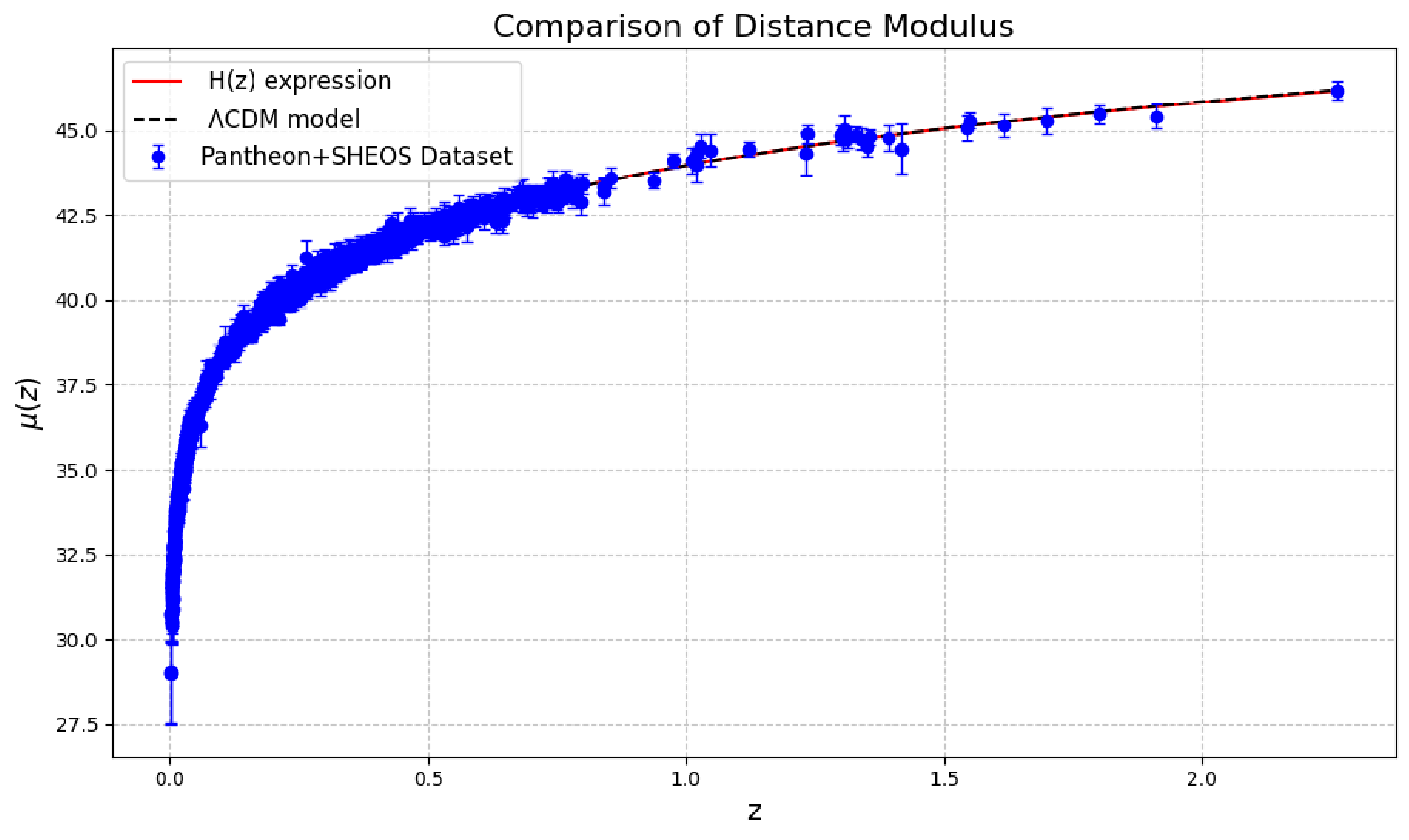}
  \caption{Error bar plots obtained from different combinations of datasets.}\label{fig:f2}
\end{figure}

To perform a comprehensive statistical analysis, we construct the total chi-square function by summing the contributions from each of the datasets described above:
\begin{equation}\label{22}
\chi^{2}_{joint}=\chi^{2}_{CC}+\chi^{2}_{BAO}+\chi^{2}_{SN}
\end{equation}
This joint methodology enables more stringent constraints on model parameters by integrating insights from diverse observational datasets. The resulting confidence contours corresponding to the Cosmic Chronometers (CC) dataset alone, as well as the combined CC+BAO and CC+BAO+Pantheon+SHOES datasets, are displayed in Figure \ref{fig:f1}. Additionally, Figure \ref{fig:f2} presents the error bar plots for individual datasets and their various combinations, visually illustrating how successive data integration refines the parameter estimates.

Using a parametrization of the deceleration parameter, our analysis of the $f(R,G)$ gravity model is resulted in consistent and physically reasonable constraints on cosmological parameters, validated across multiple datasets. Our findings indicate that the Hubble constant $H_{0}$ lies in the range $71.7–72.8$ km/s/Mpc, with a relative uncertainty of approximately $1.7\%$. Our obtained $H_{0}$ values agree with late-time observations such as SHOES $H_{0}=73.4\pm1.04$ km/s/Mpc \cite{Di21} and are somewhat larger than the Planck CMB value $H_{0}=67.4\pm0.5$ km/s/Mpc \cite{Planck20}, potentially alleviating the Hubble discrepancy. Across all dataset combinations, the present deceleration parameter, $q_{0}$ exhibits negative values, clustering around $-0.48$ to $-0.49$, affirming the Universe's ongoing accelerated expansion. These results are closely aligned with the standard cosmological model predictions ($\Lambda$CDM typically gives $q_{0}\approx-0.55$). Furthermore, the evolution parameter $q_{1}$ is found to be positive and close to unity in all cases, indicating that the rate of acceleration is increasing with time — a feature that is consistent with evolving dark energy scenarios and late-time modifications to general relativity. The constraints are reasonably well-defined, with $q_{0}$ and $q_{1}$ showing uncertainties within the $16-17\%$ range, reflecting a reliable fit to the data.

A comparative analysis is performed between our $H_{0}$ and $q_{0}$ values and those from recent modified gravity theory studies. In contrast, the Hu-Sawicki $f(R)$ gravity model yields a somewhat lower $H_{0}$ value of $69.4\pm1.6$ and $q_{0}$ of $-0.53\pm0.04$, showing better agreement with Planck data \cite{RC17}. The $f(T)$ gravity model with a power-law form yields $H_{0}=70.3\pm1.5$ and $q_{0}=-0.56 \pm 0.02$, consistent with both $\Lambda$CDM and Planck observations \cite{nesseris2013}. $f(Q)$ gravity predicts $H_{0}=71.1\pm1.8$ and $q_{0}=-0.50 \pm 0.05$ offering a moderate solution to the Hubble tension and aligning with CC+BAO+Pantheon data \cite{mandal2022}. An enhanced late-time expansion and a redshift-dependent deceleration parameter are indicated by the $f(Q,T)$ gravity reports $H_{0}=71.6\pm1.9$ and $q_{0}=-0.47$ \cite{yousaf2022}. In great agreement with SHOES local measurements, $f(T,B)$ gravity displays $H_{0}=72.3 \pm 1.7$ and $q_{0}=-0.49 \pm 0.03$ \cite{bahamonde2018}. $f(Q,C)$ gravity, supporting Barrow-type dark energy evolution, finds $H_{0}=72.5 \pm 1.3$ and $q_{0}=-0.48 \pm 0.02$, which closely aligns with our results and supports consistency with late-time cosmological observations \cite{Lobo23}. Our obtained values from the $f(R, G)$ model not only align well with the latest low-redshift observations, but also fall within the range predicted by other viable modified gravity frameworks such as $f(Q,C)$ and $f(T,B)$. Our predicted Hubble and deceleration parameter $q_{0} \approx -0.48$ to $-0.49$ support a consistent late-time accelerating Universe scenario.
\section{Analysis of cosmological parameter behavior from MCMC constraints}\label{sec4}
\hspace{0.5cm} This section explores the properties of essential cosmological parameters derived from observational constraints using MCMC methods. Based on the best-fit results and confidence intervals, we analyze the model's behavior across redshift and assess its alignment with standard cosmology. The behavior of important parameters like the deceleration parameter $q(z)$, energy density, pressure and EoS parameter $\omega(z)$ is explored in detail. The analysis illuminates the implications of the constrained model for cosmic dynamics, encompassing the transition phase and the present expansionary behavior.
\subsection{Deceleration parameter}\label{sec4.1}
\begin{figure}[hbt!]
  \centering
  \includegraphics[scale=0.4]{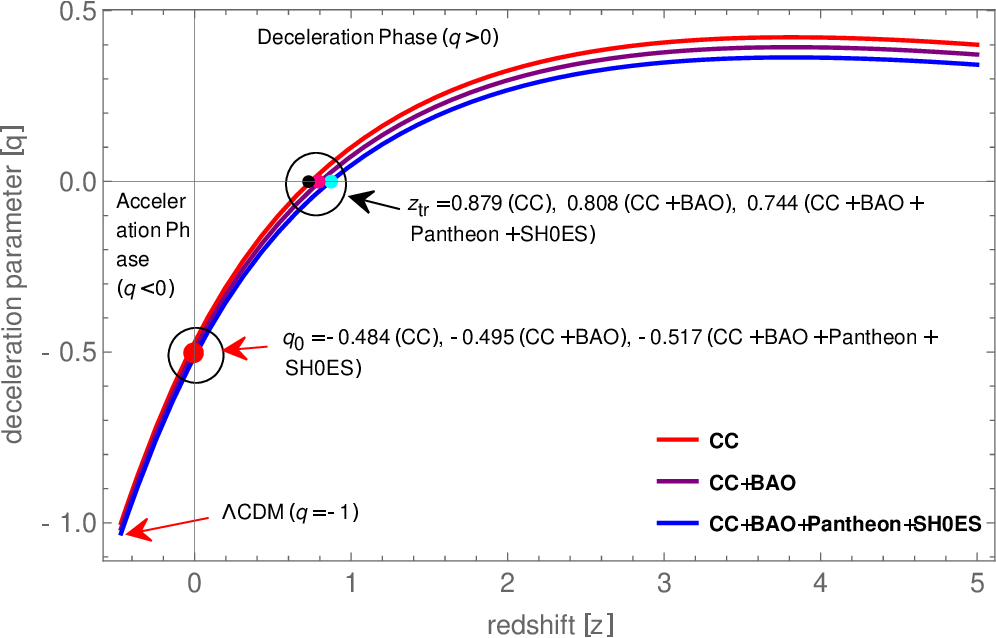}
  \caption{Redshift dependence of $q(z)$ across various observational datasets.}\label{fig:f3}
\end{figure}

\hspace{0.5cm} The optimal parameters derived from our MCMC analysis are implemented to depict the redshift dependency of the deceleration parameter $q(z)$. We display $q(z)$ in Figure \ref{fig:f3} using three dataset configurations: CC, CC+BAO and the combined CC+BAO+Pantheon+SHOES dataset.

The transition from decelerated $(q>0)$ to accelerated $(q<0)$ expansion is apparent in the plot for all three datasets. The transition redshift $z_{tr}$, marking the point where $q(z_{tr})=0$, changes depending on the dataset. Specifically, the transition occurs at: $z_{tr}=0.879$ for CC data, $z_{tr}=0.808$ for CC+BAO data and $z_{tr}=0.744$ for the full dataset (CC+BAO+Pantheon+SHOES). As more data are incorporated, $q_{0}$ becomes increasingly negative, pointing to stronger acceleration at late times. The resulting values are: $q_{0}=-0.484$ (CC), $q_{0}=-0.495$ (CC+BAO) and $q_{0}=-0.517$ (full dataset). All values are consistent with current acceleration and match the $\Lambda$CDM model's expectations of $q_{0}\approx-0.55$ to $-0.6$ \cite{Sam24}. The trend indicates that combining datasets not only reduces uncertainties but also pushes the transition to lower redshifts, further solidifying the accelerated expansion phase. The agreement between the model and observations highlights the $q(z)$ form's robustness and suitability for describing the Universe's expansion history.
\subsection{Energy density and pressure}\label{sec4.2}
\hspace{0.5cm} Our analysis now focuses on the energy density $\rho(z)$ and pressure $p(z)$ profiles in the context of our model. These variables play a central role in determining the Universe's evolution and the characteristics of the cosmic fluid. We utilize the best-fit MCMC parameters to reconstruct and analyze the redshift evolution of $\rho(z)$ and $p(z)$. Our theoretical framework yields the following forms for $\rho(z)$ and $p(z)$:
\begin{eqnarray}\label{23}
\rho(z)&=&3H_{0}^{2}(1+z)^{2+2q_{0}}e^{2q_{1}[1-cos(log(1+z))]}\bigg(1+60\alpha H_{0}^{2}(1+z)^{2+2q_{0}}e^{2q_{1}[1-cos(log(1+z))]}\big\{1+q_{0}\\\nonumber
&&+q_{1}sin(log(1+z))\big\}\bigg)-18\alpha H_{0}^{4}(1+z)^{4+4q_{0}}e^{4q_{1}[1-cos(log(1+z))]}\big(1+q_{0}+q_{1}sin(log(1+z))\big)^{2}\\\nonumber
&&-36\alpha H_{0}^{4}(1+z)^{4+4q_{0}}e^{4q_{1}[1-cos(log(1+z))]}\bigg[2(1+q_{0})\big\{1+q_{0}+q_{1}sin(log(1+z))\big\}\\\nonumber
&&+2q_{1}sin(log(1+z))\big(1+q_{0}+q_{1}sin(log(1+z))\big)+q_{1}cos(log(1+z))\bigg]-288\beta\gamma^{2}\\\nonumber
&&H_{0}^{3}(1+z)^{3+3q_{0}}e^{3q_{1}[1-cos(log(1+z))]}\bigg[-4H_{0}^{5}(1+z)^{5+5q_{0}}e^{5q_{1}[1-cos(log(1+z))]}\big(1+q_{0}\\\nonumber
&&+q_{1}sin(log(1+z))\big)+H_{0}^{3}(1+z)^{3+3q_{0}}e^{3q_{1}[1-cos(log(1+z))]}\bigg\{2(1+q_{0})\big(1+q_{0}\\\nonumber
&&+q_{1}sin(log(1+z))\big)+2q_{1}sin(log(1+z))\big(1+q_{0}+q_{1}sin(log(1+z))\big)+q_{1}cos(log(1+z))\bigg\}\bigg]\\\nonumber
&&-\frac{\beta}{2}exp\bigg[-24\gamma H_{0}^{4}(1+z)^{4+4q_{0}}e^{4q_{1}[1-cos(log(1+z))]}(q_{0}+q_{1}sin(log(1+z)))\bigg]\\\nonumber
&&\bigg\{-24\gamma H_{0}^{4}(1+z)^{4+4q_{0}}e^{4q_{1}[1-cos(log(1+z))]}(q_{0}+q_{1}sin(log(1+z)))-1\bigg\},
\end{eqnarray}
\begin{figure}[hbt!]
  \centering
  \includegraphics[scale=0.4]{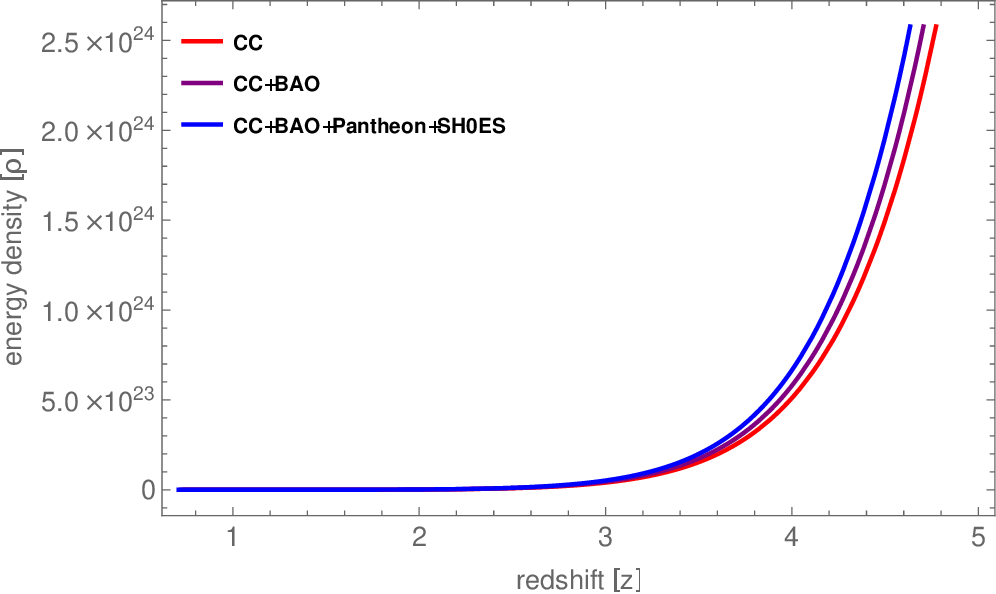}~~~
  \includegraphics[scale=0.4]{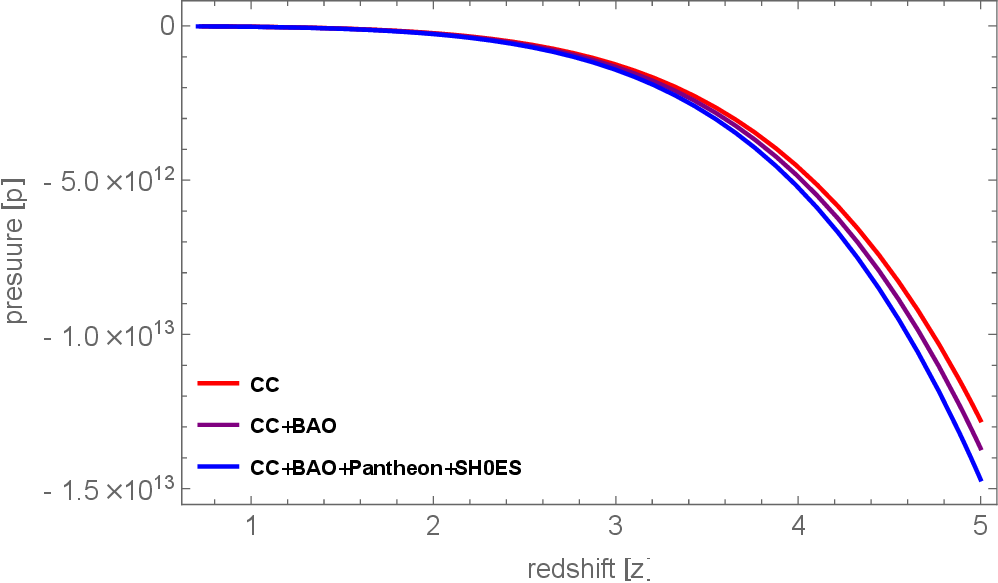}
  \caption{Redshift dependence of $\rho(z)$ and $p(z)$ for the parameters $\alpha=0.5$, $\beta=0.6$ and $\gamma=0.7$.}\label{fig:f4}
\end{figure}
\begin{eqnarray}\label{24}
p(z)&=&-H_{0}^{2}(1+z)^{2+2q_{0}}e^{2q_{1}[1-cos(log(1+z))]}\big(1-2q_{0}-2q_{1}sin(log(1+z))\big)-54\alpha H_{0}^{4}\\\nonumber
&&(1+z)^{4+4q_{0}}e^{4q_{1}[1-cos(log(1+z))]}\big(1+q_{0}+q_{1}sin(log(1+z))\big)^{2}+12\alpha H_{0}^{3}(1+z)^{3+3q_{0}}\\\nonumber
&&e^{3q_{1}[1-cos(log(1+z))]}\big(1+q_{0}+q_{1}sin(log(1+z))\big)-\frac{\beta}{2}exp\bigg[-24\gamma H_{0}^{4}(1+z)^{4+4q_{0}}\\\nonumber
&&e^{4q_{1}[1-cos(log(1+z))]}(q_{0}+q_{1}sin(log(1+z)))\bigg]\bigg\{-24\gamma H_{0}^{4}(1+z)^{4+4q_{0}}\\\nonumber
&&e^{4q_{1}[1-cos(log(1+z))]}(q_{0}+q_{1}sin(log(1+z)))-1\bigg\}-24\alpha H_{0}^{4}(1+z)^{4+4q_{0}}\\\nonumber
&&e^{4q_{1}[1-cos(log(1+z))]}\bigg[2(1+q_{0})\big\{1+q_{0}+q_{1}sin(log(1+z))\big\}+2q_{1}sin(log(1+z))\\\nonumber
&&\big(1+q_{0}+q_{1}sin(log(1+z))\big)+q_{1}cos(log(1+z))\bigg]-12\alpha H_{0}^{4}(1+z)^{4+4q_{0}}\\\nonumber
&&e^{4q_{1}[1-cos(log(1+z))]}\bigg[(1-3q_{0}-3q_{1}sin(log(1+z))\bigg\{2\big(1+q_{0}+q_{1}sin(log(1+z))\big)^{2}\\\nonumber
&&+q_{1}cos(log(1+z))\bigg\}-4(1+q_{0})q_{1}cos(log(1+z))-2q_{1}^{2}sin[2(log(1+z))]+\\\nonumber
&&q_{1}sin(log(1+z))\bigg]-32\beta\gamma^{2}H_{0}^{3}(1+z)^{3+3q_{0}}e^{3q_{1}[1-cos(log(1+z))]}\bigg[-q_{0}\\\nonumber
&&-q_{1}sin(log(1+z))3H_{0}(1+z)^{1+q_{0}}e^{q_{1}[1-cos(log(1+z))]}\bigg\{2\big(1+q_{0}+q_{1}sin(log(1+z))\big)^{2}\\\nonumber
&&+q_{1}cos(log(1+z))\bigg\}\bigg]-4\beta\gamma^{2}exp\bigg[-24\gamma H_{0}^{4}(1+z)^{4+4q_{0}}e^{4q_{1}[1-cos(log(1+z))]}\\\nonumber
&&(q_{0}+q_{1}sin(log(1+z)))\bigg]H_{0}^{2}(1+z)^{2+2q_{0}}e^{2q_{1}[1-cos(log(1+z))]}\bigg\{576\gamma\bigg[16H_{0}^{10}(1+z)^{10+10q_{0}}\\\nonumber
&&e^{10q_{1}[1-cos(log(1+z))]}\big(1+q_{0}+q_{1}sin(log(1+z))\big)^{2}+H_{0}^{6}(1+z)^{6+6q_{0}}e^{6q_{1}[1-cos(log(1+z))]}\\\nonumber
&&\bigg(2\big(1+q_{0}+q_{1}sin(log(1+z))\big)^{2}+q_{1}cos(log(1+z))\bigg)^{2}-H_{0}^{8}(1+z)^{8+8q_{0}}\\\nonumber
&&e^{8q_{1}[1-cos(log(1+z))]}\bigg(\big\{1+q_{0}+q_{1}sin(log(1+z))\big\}^{2}\bigg[2\big(1+q_{0}+q_{1}sin(log(1+z))\big)^{2}\\\nonumber
&&+q_{1}cos(log(1+z))\bigg]\bigg)\bigg]+96H_{0}^{6}(1+z)^{6+6q_{0}}e^{6q_{1}[1-cos(log(1+z))]}\bigg[5\big(1+q_{0}+q_{1}sin(log(1+z))\big)^{2}\\\nonumber
&&+q_{1}cos(log(1+z))\bigg]-24H_{0}^{4}(1+z)^{4+4q_{0}}e^{4q_{1}[1-cos(log(1+z))]}\bigg[3\big(1+q_{0}+q_{1}sin(log(1+z))\big)\\\nonumber
&&\bigg(2\big(1+q_{0}+q_{1}sin(log(1+z))\big)^{2}+q_{1}cos(log(1+z))\bigg)+4(1+q_{0})q_{1}cos(log(1+z))\\\nonumber
&&+2q_{1}^{2}sin[2(log(1+z))]-q_{1}sin(log(1+z))\bigg]\bigg\}
\end{eqnarray}

In Figure \ref{fig:f4}, we plot the $\rho(z)$ and $p(z)$ against redshift. The energy density $\rho(z)$ exhibits a monotonic increase with redshift, which is consistent with the Universe's early matter and radiation dominance. Throughout the redshift range, $p(z)$ is consistently negative, consistent with dark energy's influence on the Universe's accelerating expansion. The overlap of curves for all datasets shows the model's stability: combination of BAO, Pantheon and SHOES data improves accuracy without altering $\rho(z)$ and $p(z)$ trends. This agreement suggests a robust and coherent picture of the Universe's evolution. Our analysis reinforces the $\Lambda$CDM-like paradigm, where dark energy acts like a cosmological constant and the model's stability under varied observational constraints.
\subsection{EoS parameter and its evolution}\label{sec4.3}
\hspace{0.5cm} We explore the equation of state parameter $\omega(z)$, quantifying the relationship between pressure and energy density as $\omega(z)=\frac{p(z)}{\rho(z)}$. By studying $\omega(z)$, we gain insight into dark energy's characteristics and its potential departure from the cosmological constant model. Based on the reconstructed $\rho(z)$ and $p(z)$, we compute and display the EoS parameter for various dataset combinations. This analysis reveals whether the cosmic fluid's dynamics are driven by quintessence, phantom energy, or a cosmological constant. The expression for EoS parameter is given by:
\begin{equation}\label{25}
\omega=-1-\frac{2\dot{H}f_{R}-H\dot{f}_{R}+\ddot{f}_{R}+4H^{2}\ddot{f}_{G}-4H^{3}\dot{f}_{G}+8H\dot{H}\dot{f}_{G}}{3H^{2}f_{R}-\bigg(\frac{Rf_{R}+Gf_{G}-f(R,G)}{2}\bigg)+3H\dot{f}_{R}+12H^{3}\dot{f}_{G}},
\end{equation}
\begin{figure}[hbt!]
  \centering
  \includegraphics[scale=0.4]{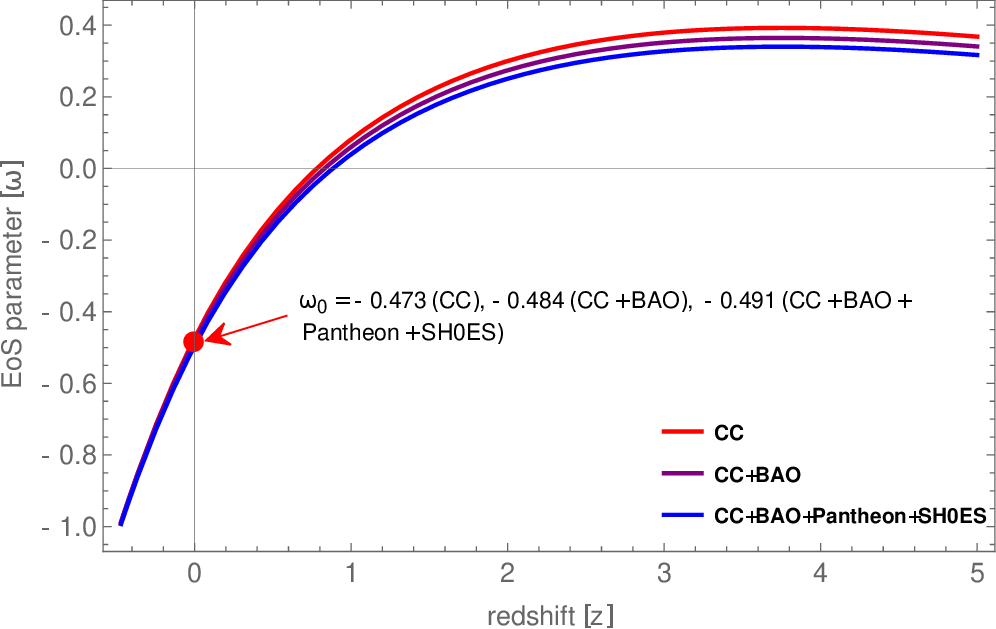}
  \caption{Redshift dependence of $\omega(z)$ for the parameter set $\alpha=0.5$, $\beta=0.6$ and $\gamma=0.7$.}\label{fig:f5}
\end{figure}

Figure \ref{fig:f5} displays the evolution of the equation of state (EoS) parameter $\omega(z)$ with redshift $z$. At high redshift $(z>>1)$, the EoS parameter $\omega(z)\approx0.33$, which is characteristic of a radiation-dominated Universe. This indicates that the model correctly reproduces the expected early-time cosmological behavior, where radiation dominates the energy content and the influence of dark energy is negligible. As the redshift decreases (meaning the Universe evolves toward the present) the value of $\omega(z)$ decreases smoothly, becoming more negative. This trend reflects the gradual emergence and increasing influence of the dark energy component. Around redshift $z\approx0.5-1$, dark energy begins to dominate the cosmic expansion, leading to the onset of accelerated expansion. The present-day values of $\omega(z)$ are: $\omega_{0}=-0.473$, $-0.484$ and $-0.491$ for the CC, CC+BAO and CC+BAO+Pantheon+SHOES, respectively. The present-day $\omega$ values indicate a quintessence-like behavior for dark energy, but the model's proximity to $-1$ ensures that it remains observationally viable. Incorporating BAO, Pantheon and SHOES data further constrains the EoS estimate, strengthening the viability of a dynamical dark energy model.
\section{Analysis of energy conditions}\label{sec5}
\hspace{0.5cm} To further examine the physical viability of our model, we investigate the standard energy conditions, namely the Null Energy Condition (NEC), the Dominant Energy Condition (DEC) and the Strong Energy Condition (SEC). These conditions serve as theoretical constraints on the energy-momentum tensor and are widely used to test the consistency of gravitational models. Given the reconstructed energy density $\rho(z)$ and pressure $p(z)$, the energy conditions can be expressed as:
\begin{equation}\label{26}
\text{NEC:}\hspace{0.3cm} \rho+p=H\dot{f}_{R}-2\dot{H}f_{R}+4H^{3}\dot{f}_{G} -8H\dot{H}\dot{f}_{G}-4H^{2}\ddot{f}_{G}\geq 0,
\end{equation}
\begin{eqnarray}\label{27}
\text{DEC:}\hspace{0.3cm} \rho-p&=&6H^{2}f_{R}+2\dot{H}f_{R}-\bigg(Rf_{R}+Gf_{G}-f(R,G)\bigg)+5H\dot{f}_{R}+20H^{3}\dot{f}_{G}+\ddot{f}_{R}\\\nonumber
&&+8H\dot{H}\dot{f}_{G}+4H^{2}\ddot{f}_{G}\geq 0,
\end{eqnarray}
\begin{eqnarray}\label{28}
\text{SEC:}\hspace{0.3cm} \rho+3p&=&-6H^{2}f_{R}-6\dot{H}f_{R}+\bigg(Rf_{R}+Gf_{G}-f(R,G)\bigg)-3H\dot{f}_{R}-3\ddot{f}_{R}-12H^{2}\ddot{f}_{G}\\\nonumber
&&-12H^{3}\dot{f}_{G}-24H\dot{H}\dot{f}_{G} \geq 0.
\end{eqnarray}
\begin{figure}
  \centering
  \includegraphics[scale=0.4]{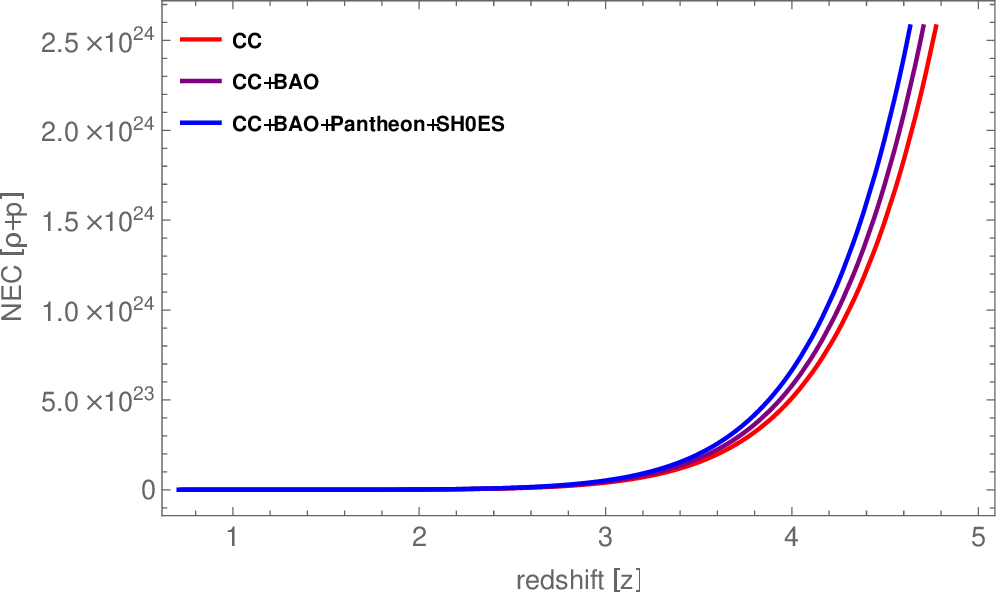}~
  \includegraphics[scale=0.4]{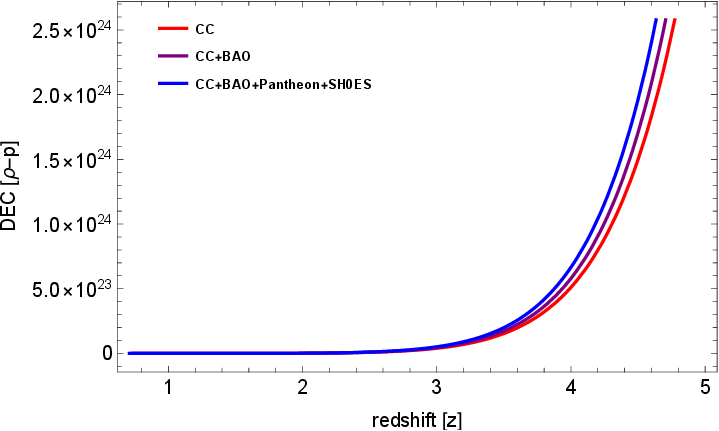}\\
  \includegraphics[scale=0.4]{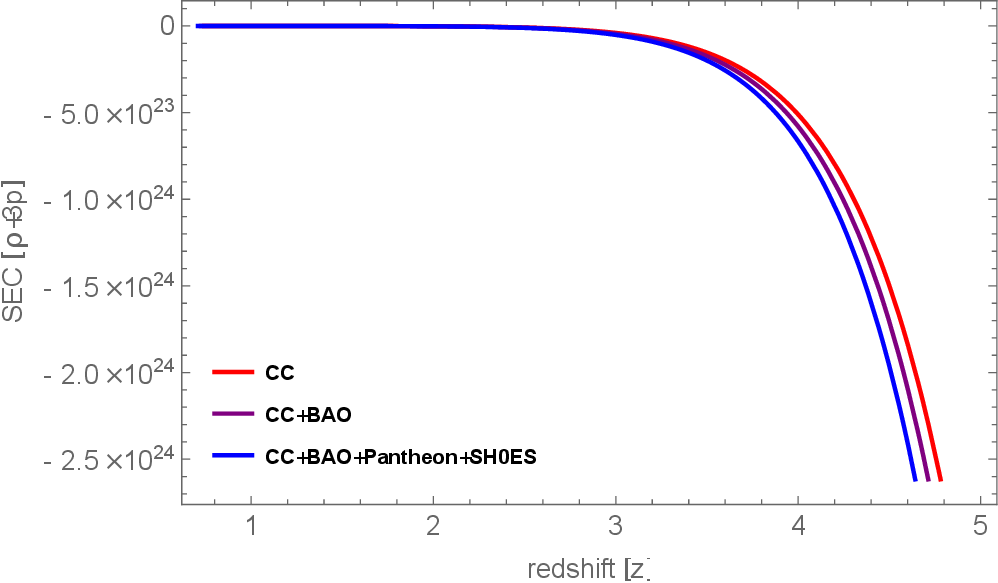}
  \caption{Redshift evolution of energy conditions for the parameter set $\alpha=0.5$, $\beta=0.6$ and $\gamma=0.7$.}\label{fig:f6}
\end{figure}

Figure \ref{fig:f6} demonstrates the energy condition's variation with redshift, showcasing its adherence or deviation throughout the Universe's evolution. Our results indicate that both the NEC and DEC are satisfied throughout the cosmic evolution, ensuring that the energy density is non-negative and that the flow of energy respects causality. These conditions also imply that the effective cosmic fluid behaves in a physically acceptable manner. On the other hand, the SEC is found to be violated at late times, which aligns with the requirement for cosmic acceleration. The violation of SEC supports the presence of a repulsive gravitational component (i.e., dark energy) responsible for the observed late-time accelerated expansion of the Universe. This behavior is consistent with current cosmological observations and reinforces the validity of our model \cite{Amit24}.
\section{Statefinder diagnostic analysis}\label{sec6}
\hspace{0.5cm} To further explore the nature of dark energy in our model, we employ the statefinder diagnostic introduced by \cite{Sahni03}, which provides a geometrical method to differentiate between various cosmological models by using higher derivatives of the scale factor. The statefinder parameters $\{r, s\}$ are defined as:
\begin{equation}\label{29}
  r=\frac{\dddot a}{aH^{3}}=2q^{2}+q+(1+z)\frac{dq}{dz},
\end{equation}
\begin{equation}\label{30}
  s=\frac{(r-1)}{3(q-\frac{1}{2})}. \bigg(q\neq\frac{1}{2}\bigg).
\end{equation}
These parameters are particularly useful because the $\Lambda$CDM model corresponds to a fixed point $\{r, s\}=(1, 0)$. Any deviation from this point indicates departure from the standard cosmological model, thus providing a diagnostic criterion for comparing the dynamic behavior of dark energy models. Using the form of the deceleration parameter and the Hubble function in our model, the expressions for the statefinder parameters $\{r, s\}$ can be explicitly written as:
\begin{equation}\label{31}
r=2\big[q_{0}+q_{1}sin\big(log[1+z]\big)\big]^{2}+q_{0}+q_{1}sin(log[1+z])+q_{1}cos(log[1+z]),
\end{equation}
\begin{equation}\label{32}
s=\frac{2\big[q_{0}+q_{1}sin\big(log[1+z]\big)\big]^{2}+q_{0}+q_{1}sin(log[1+z])+q_{1}cos(log[1+z])}{3\bigg[q_{0}+q_{1}sin(log[1+z])-\frac{1}{2}\bigg]}.
\end{equation}
\begin{figure}[hbt!]
  \centering
  \includegraphics[scale=0.4]{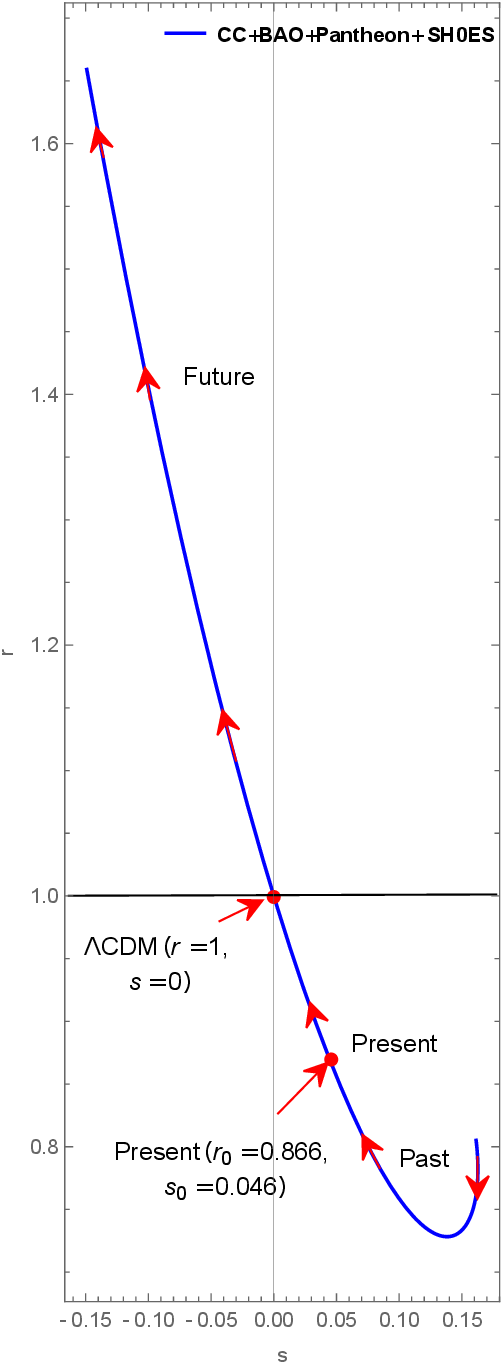}
  \caption{Statefinder trajectory in the $\{r, s\}$ plane for the CC+BAO+Pantheon+SHOES dataset.}\label{fig:f7}
\end{figure}

The $\{r, s\}$ trajectory shown in Figure \ref{fig:f7} provides significant insights into the nature of the cosmic expansion in our model. The present-day values of the statefinder parameters are $r_{0}=0.866$ and $s_{0}=0.046$, which lie in close proximity to the $\Lambda$CDM point $(1, 0)$ \cite{Alam2024}. The trajectory evolves from the region corresponding to earlier cosmic epochs (larger $r$, positive $s$) and moves toward the $\Lambda$CDM point in the future, demonstrating a transition from quintessence behavior to a phase closely resembling it. The fact that $r < 1$ and $s > 0$ at present suggests a dark energy model that is slightly dynamical, differing from a pure cosmological constant but still consistent with an accelerating Universe. These findings support the interpretation that the model behaves in a $\Lambda$CDM-like manner at low redshifts, further strengthening its compatibility with current cosmological observations.
\section{Estimation of the age of the Universe}\label{sec7}
\hspace{0.5cm} Verifying the Universe's age is vital for ensuring the accuracy of cosmological models. This can be achieved using the expression
\begin{equation}\label{33}
t_{0}-t=\int_{0}^{z}\frac{dz}{(1+z)H(z)},
\end{equation}
Using our model's Hubble parameter (\ref{13}), we calculate the time evolution function $H_{0}(t_{0}-t)$ in conjunction with this result,
\begin{equation}\label{34}
H_{0}(t_{0}-t)=\int_{0}^{z}\frac{dz}{(1+z)^{2+2q_{0}}exp\big[q_{1}\{1-cos(log(1+z))\}\big]}.
\end{equation}
\begin{figure}[hbt!]
  \centering
  \includegraphics[scale=0.4]{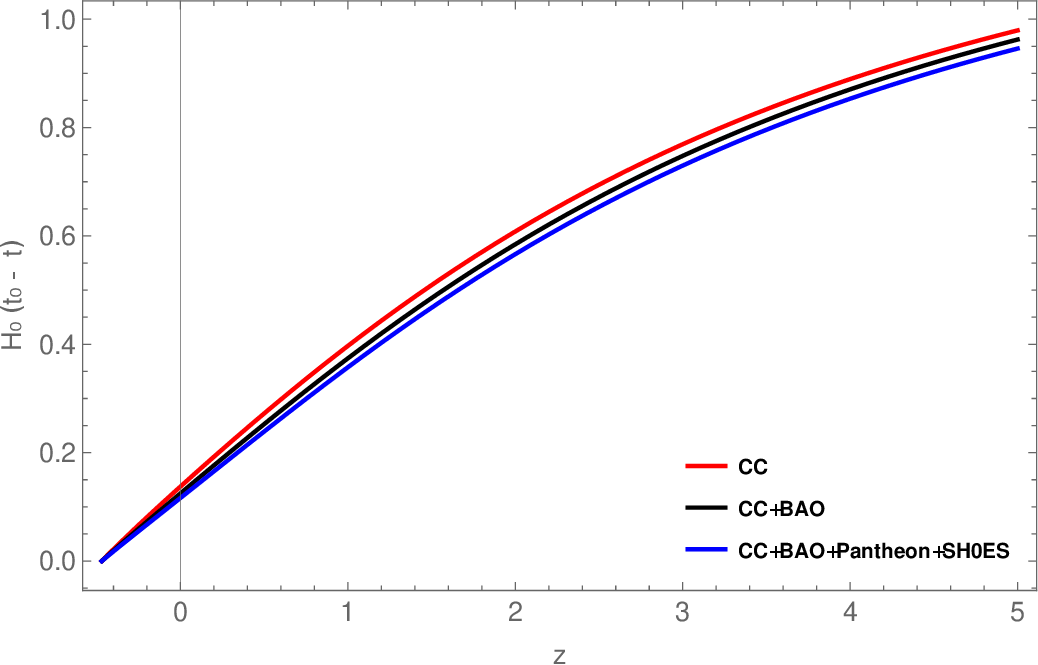}
  \caption{The variation of cosmic time with redshift for best fit values.}\label{fig:f8}
\end{figure}

Figure \ref{fig:f8} emphasises the graphic visualisation of the cosmic time-redshift relationship. In the high-redshift limit, $H_{0}(t_{0}-t)$ asymptotes to a fixed value $H_{0}t_{0}$, representing the Universe's age in Hubble units. For our model, the asymptotic values of this function—i.e. $H_{0}t_{0}$—are: $0.975$ for the CC dataset, $0.961$ for the CC+BAO dataset and $0.945$ for the CC+BAO+Pantheon+SHOES dataset. These translate into Universe ages $(t_{0})$ of approximately: $13.59$ billion years (CC), $13.34$ billion years (CC+BAO) and $13.01$ billion years (CC+BAO+Pantheon+SHOES). Our model's cosmic age estimates align with recent observations, demonstrating its ability to make accurate predictions and reinforcing its validity in late-time cosmology \cite{Cruz20,IB20}.
\section{Thermodynamic consistency and entropy evolution in the expanding Universe}\label{sec8}
\hspace{0.5cm} By applying the laws of thermodynamics to the $f(R,G)$ gravity framework, we investigate the Universe's thermodynamic properties, commencing with the first law for an ideal fluid in a volume $V$ \cite{V16}.
\begin{equation}\label{35}
Tds=d(V\rho)+pdV,
\end{equation}
We reformulate it as:
\begin{equation}\label{36}
Tds=d((\rho+p)V)-Vdp,
\end{equation}
and by using the thermodynamic identity:
\begin{equation}\label{37}
dp=\bigg[\frac{\rho+p}{T}\bigg]dT.
\end{equation}
We arrive at the entropy $(s)$ differential:
\begin{equation}\label{38}
ds=\frac{d((\rho+p)V)}{T}-(\rho+p)V\frac{dT}{T^{2}},
\end{equation}
which simplifies to:
\begin{equation}\label{39}
ds=d\bigg(\frac{(\rho+p)V}{T}\bigg),
\end{equation}
Upon performing the integration, the total entropy expression becomes:
\begin{equation}\label{40}
s=\frac{(\rho+p)V}{T},
\end{equation}
For the entropy density $(\mathbb{S})$, we define:
\begin{equation}\label{41}
\mathbb{S}=\frac{s}{V}=\frac{(\rho+p)}{T}=\frac{(1+\omega)\rho}{T}.
\end{equation}
Assuming a barotropic fluid with $p=\omega\rho$, where $0<\omega<1$. Using this, the first law can be expressed as:
\begin{equation}\label{42}
d(V\rho)+\omega\rho dV=(1+\omega)Td\bigg[\frac{V\rho}{T}\bigg],
\end{equation}
Integrating the differential form:
\begin{equation}\label{43}
\omega d\rho=(1+\omega)\rho\frac{dT}{T},
\end{equation}
yields the temperature as a function of energy density:
\begin{equation}\label{44}
T=\rho^{\frac{\omega}{1+\omega}}=\bigg[3H^{2}f_{R}-\bigg(\frac{Rf_{R}+Gf_{G}-f(R,G)}{2}\bigg)+3H\dot{f}_{R}+12H^{3}\dot{f}_{G}\bigg]^{\frac{\omega}{1+\omega}},
\end{equation}
Substituting this back gives the entropy density as:
\begin{equation}\label{66}
\mathbb{S}=(1+\omega)\rho^{\frac{1}{1+\omega}}=(1+\omega)\bigg[3H^{2}f_{R}-\bigg(\frac{Rf_{R}+Gf_{G}-f(R,G)}{2}\bigg)+3H\dot{f}_{R}+12H^{3}\dot{f}_{G}\bigg]^{\frac{1}{1+\omega}}.
\end{equation}
\begin{figure}[hbt!]
  \centering
  \includegraphics[scale=0.4]{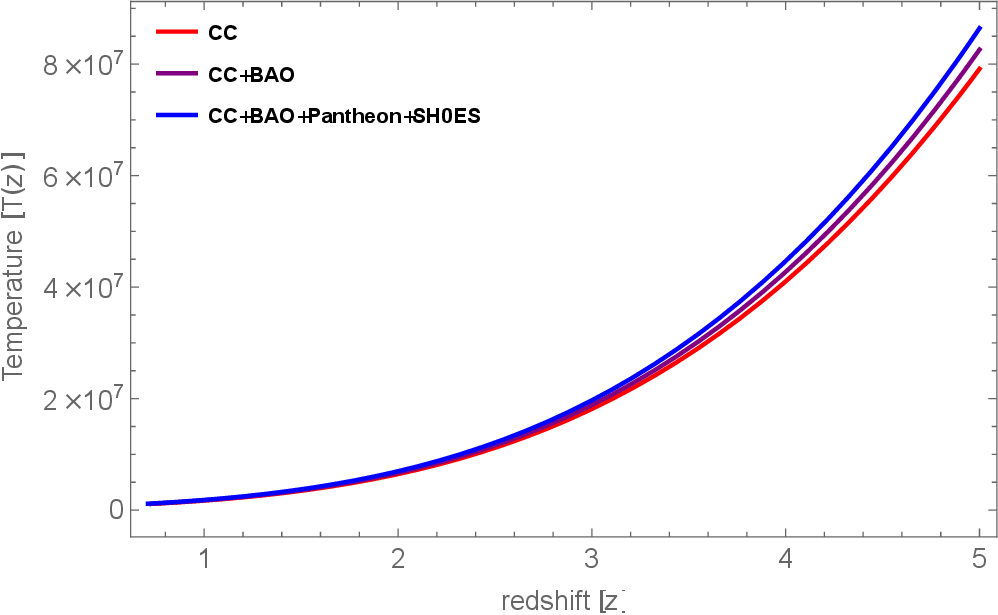}~
  \includegraphics[scale=0.4]{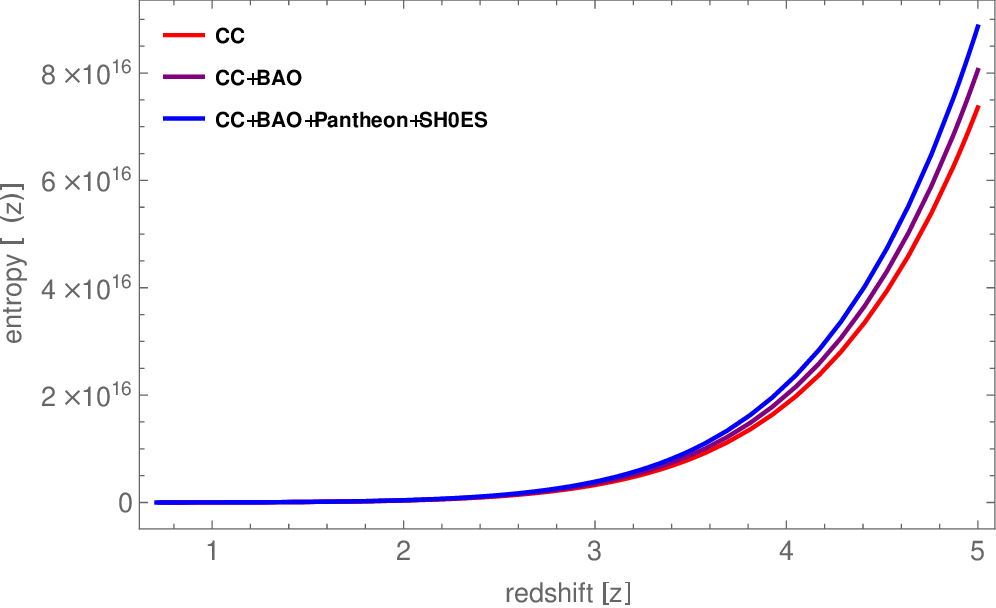}
  \caption{Redshift evolution of temperature and entropy density for the parameter set $\alpha=0.5$, $\beta=0.6$ and $\gamma=0.7$.}\label{fig:f9}
\end{figure}

The temperature and entropy density profiles of our $f(R,G)$ model, shown in Figure \ref{fig:f9}, demonstrate their redshift dependence under various dataset constraints. The temperature curve in Figure \ref{fig:f9} reveals that the Universe was significantly hotter in the past, with temperature increasing towards the higher redshifts. The model's thermal behavior aligns with established cosmological principles, confirming its ability to describe the Universe's expected thermal evolution. Towards the future $(z\rightarrow-1)$, the temperature asymptotically approaches to a constant, indicating thermal stabilization in a low-energy phase. The entropy density, shown in Figure \ref{fig:f9}, grows with redshift, reflecting higher entropy content in earlier epochs. This finding matches the expected trend of increasing energy density and temperature in the Universe's early epochs. The expanding Universe leads to a gradual decrease in entropy density, but total entropy still rises, adhering to the generalized second law. Our derived entropy density form $\mathbb{S}\propto\rho^{\frac{1}{1+\omega}}$ reveals a strong interplay between thermodynamics and energy density dynamics. These findings not only confirm the internal consistency of the thermodynamic framework but also support the broader physical acceptability of the $f(R,G)$ gravity theory as a viable candidate for describing cosmic evolution.
\section{Conclusion}\label{sec9}
\hspace{0.5cm} Our work focuses on the cosmological aspects of a modified gravity theory within the $f(R,G)$ framework, adopting the specific form $f(R,G)=R+\alpha R^{2}+\beta e^{\gamma G}$. By adopting a logarithmic modulation for the deceleration parameter $q(z)=q_{0}+q_{1}sin[log(1+z)]$, we analytically determined $H(z)$ and constrained the parameters using MCMC simulations with CC, BAO and Pantheon+SHOES datasets.

Our MCMC analysis yielded consistent and physically meaningful constraints. We found the Hubble constant $H_{0}$ to range from $71.7$ to $72.8$ km/s/Mpc, with $1.7\%$ relative uncertainty, showing good agreement with late-time observations like SHOES and slightly exceeding the Planck CMB values. Our model appears to offer a potential solution to the Hubble tension problem. The deceleration parameter $q_{0}$ is firmly established in the negative range, around $–0.48$ to $–0.49$, confirming cosmic acceleration. Furthermore, the evolution parameter $q_{1}$, found to be positive and close to unity, indicates that the acceleration is increasing with time — a feature compatible with evolving dark energy or modified gravity scenarios. We determined the transition redshift $z_{tr}$ from the behavior of the deceleration parameter $q(z)$, indicating when the Universe switched from deceleration to acceleration. The transition from deceleration to acceleration happens earlier $(z_{tr}= 0.879)$ with CC data alone, but later $(z_{tr}= 0.744)$ with the CC+BAO+Pantheon+SHOES dataset, pointing to stronger acceleration at late times with more stringent constraints.

We also reconstructed the energy density $\rho(z)$ and pressure $p(z)$, which displayed physically consistent behavior: $\rho(z)$ increases monotonically with redshift, reflecting radiation and matter dominance at early times, while $p(z)$ remains negative, supporting the presence of a repulsive dark energy component. The EoS parameter $\omega(z)$ evolves from a radiation-like behavior $(\omega\approx0.33)$ at high redshift to increasingly negative values at low redshift, with present-day values between $-0.473$ and $-0.491$ depending on the dataset. This transition aligns well with the standard cosmological picture of dark energy dominance at late times.

Our examination reveals that NEC and DEC are satisfied, while SEC violation at late times supports cosmic acceleration, consistent with expectations from GR and modified gravity models. Further insights were obtained through the statefinder diagnostic. The trajectory in the $\{r,s\}$ plane shows that our model evolves toward the $\Lambda$CDM fixed point $\{r=1,\;s=0\}$. The present-day values $r_{0}=0.866$ and $s_{0}=0.046$ lie close to this point.

Based on our Hubble function reconstruction, the age parameter $H_{0}t_{0}$ was found to be $0.975$, $0.961$ and $0.945$ for the CC, CC+BAO and joint datasets, respectively. These correspond to cosmic ages of approximately $13.59$, $13.34$ and $13.01$ billion years which is in agreement with current observational bounds, further supporting the viability of our model.

Finally, the thermodynamic properties of the model were analyzed. The temperature evolution $T(z)$ rises with redshift, consistent with a hot early Universe and cooling as it expands. The entropy density $\mathbb{S}(z)$ increases toward higher redshifts, mirroring the rise in energy content and gradually decreases at low redshift. Importantly, total entropy increases over time, confirming that the generalized second law of thermodynamics is expected within our model.

In summary, the $f(R,G)$ gravity model with a logarithmic deceleration parameter provides a viable and robust description of the Universe's expansion history. It is consistent with key observational datasets, captures the transition from deceleration to acceleration, supports thermal and energy condition criteria and offers promising features that may help to resolve existing cosmological tensions.

 \end{document}